\documentclass[twocolumn,aps,prl,showpacs]{revtex4}
\usepackage{graphicx}
\usepackage{epsf}
\usepackage{amsmath}
\usepackage{amssymb}

\begin{document}

\title{Odd Parity and Line Nodes in Non-Symmorphic Superconductors}

\author{T. Micklitz} 

\affiliation{Materials Science Division, Argonne National Laboratory,
  Argonne, Illinois 60439}

\author{M. R. Norman} 

\affiliation{Materials Science Division, Argonne National Laboratory,
  Argonne, Illinois 60439}

\date{\today} 


\begin{abstract}

Group theory arguments have been invoked  to argue that odd parity 
order parameters cannot have line nodes in the presence of spin-orbit coupling.
In this paper we show that these arguments 
do not hold for certain non-symmorphic superconductors. Specifically, we demonstrate 
that when the underlying crystal has a twofold screw axis, half of the odd parity 
representations vanish on the Brillouin zone face perpendicular to this axis. 
Many unconventional superconductors 
have non-symmorphic space groups, 
and we discuss implications for several materials, including UPt$_3$,
UBe$_{13}$, Li$_2$Pt$_3$B and Na$_4$Ir$_3$O$_8$.

\end{abstract}
\maketitle

Unconventional superconducting materials
include heavy fermion metals~\cite{heavyfermion}, organics~\cite{organic}, and
cuprates~\cite{hightc}.
 The unconventionality
of these materials is reflected in the symmetry of the 
Cooper pair wavefunction:
in contrast to their `conventional' counterparts, 
unconventional Cooper pairing not only breaks
gauge but also crystal symmetry. This opens the 
possibility of odd parity pairing where by fermion antisymmetry, the spins 
are in a triplet state.

Among  unconventional superconductors, an important class are 
those whose order parameter vanishes somewhere on the Fermi surface.
The presence or absence of these nodes determines the
low energy excitations, and thus the
low temperature thermodynamic and transport properties. 
It is generally stated that in the presence of spin-orbit interactions,
an odd parity order parameter cannot have a line of nodes 
on the Fermi surface.  This is known as Blount's theorem~\cite{blount}. In contrast, this
restriction does not exist for an even parity order parameter.
There are, though, several heavy fermion superconductors where Knight shift
data indicate that the Cooper pair spins are in a triplet state, yet thermodynamic
measurements imply the existence of a line of nodes~\cite{tou}.

The aim of the present paper is to investigate the generality 
of Blount's arguments.
Specifically, we show that in crystals with a twofold screw axis, 
line nodes are possible whenever the Fermi surface intersects the
Brillouin zone face perpendicular to this axis, even
in the presence of spin-orbit interactions. 
Since many unconventional
superconductors have non-symmorphic space groups, it
provides a large class of counterexamples to Blount's theorem.
We discuss implications for several materials of interest.

In the presence of spin-orbit coupling, spin is no longer a good quantum
number.  On the other hand, Anderson has shown that because of fermion
antisymmetry, one can write down analogues of Cooper pair singlets and triplets~\cite{anderson}.
By Kramers theorem, one has two degenerate states present at {\bf k}.  Coupling
them to the two degenerate states at {\bf-k}, one has an even parity state that
is a `pseudo-spin' singlet, and an odd parity state that is a `pseudo-spin' triplet.
Blount has shown, though, that this puts restrictions on the form of the odd parity state~\cite{blount}.
A node requires 
the simultaneous
fulfilling of two real equations.  Since two equations in three
variables are commonly satisfied on curves,
and these intersect the Fermi surface at isolated points, nodes for the odd parity 
Cooper pair wavefunction 
should only occur for
points on the Fermi surface.
To assure that symmetry cannot force an increase in the 
size of the nodal regions, Blount discusses the presence 
of mirror planes. He argues that 
pseudo-spin components of the odd-parity 
wavefunction form an axial vector, 
whose components parallel and
perpendicular to the plane transform 
according to different representations.
Symmetry may only force one of these component to vanish,
and therefore a larger region of zero gap is
`vanishingly improbable'.

Blount's symmetry considerations obviously apply to point group 
operations and, consequently, to any symmorphic space group. 
A {\it non}-symmorphic space group, on the other hand, contains 
screw axes and glide planes, i.e. the combined operation of
point group elements with non-primitive translations.
The latter generate additional phase factors,
which in special situations may conspire in a way that 
{\it all} of the order parameter's pseudo-spin components transform 
according to the {\it  same} representation~\cite{example}. 
In this case, symmetry
enforces the vanishing of the order parameter belonging 
to some representations.
A particular example 
of such a case was encountered
for the hexagonal close packed lattice of UPt$_3$, where
this was shown explicitly by
construction from the single electron wavefunctions~\cite{mike}.
That is, for odd parity representations that are also odd under the 
symmetry operation $z\rightarrow -z$, it was claimed that
all pseudo-spin components vanish on the hexagonal zone face
$k_z=\pi/c$.
We will employ group theory arguments to illustrate the generality
of this argument.

{\it Group Theory:---}The group theory approach to 
Cooper pair wavefunctions of unconventional superconductors
goes back to Anderson~\cite{anderson}. Classifications
of pair states at the zone center rely on irreducible 
representations of point groups and have been listed for 
many relevant crystal symmetries~\cite{G-classification}. 
Building on work on anti-symmetrized Kronecker squares 
of induced representations~\cite{bradely},  a more general space group approach
has been developed to deduce 
which Cooper pair symmetries are allowed
at arbitrary points in the Brillouin zone~\cite{yarzhemsky}.

Here, we consider a general
non-symmorphic space group symmetry $G$,
containing inversion symmetry $(I,0)$ 
and a twofold screw axis 
$(2_z, \mathbf{t}_2)$. The latter is
a symmetry operation combining
a $\pi$ rotation $2_z$ around some axis
with a non-primitive translation 
$\mathbf{t}_2=\frac{c}{2}\mathbf{e}_z$ along 
the axis by half of the lattice displacement $c$ 
(we choose this axis to define the $z$ direction). 
$(E,0)$ denotes the identity element
and $\sigma_z=I 2_z$. Also, we assume
the presence of spin-orbit interactions.

The space group approach calculates
representations ${\cal P}(\mathbf{k})$ of the Cooper pair wavefunction  
at a given $\mathbf{k}$ point in the Brillouin zone 
by the method of induced representations~\cite{bradely,brad72,yarzhemsky}, 
\begin{align}
\label{eq:1}
{\cal P}(\mathbf{k}) = \sum_\sigma P^-_\sigma \uparrow G.
\end{align}
To explain this notation,
we proceed in two steps: (i) we 
outline the construction of representations $P^-_\sigma$ 
and (ii) we indicate a prescription how 
induced representations $P^-_\sigma \uparrow G$
may be calculated. 

Concerning (i), representations $P^-_\sigma$ at a $\mathbf{k}$ point 
in the zone are constructed from small 
representations $\gamma_\mathbf{k}$ of the symmetry group
of wavevector $\mathbf{k}$ (`little group' $G^\mathbf{k}$). 
Referring for details to 
Refs.~\onlinecite{yarzhemsky,bradely,brad72}
we merely state the procedure:  the sum $\sigma$ 
in Eq.~(\ref{eq:1}) extends over   
those representatives $d_\sigma$ in a double coset decomposition 
$G = \sum_\alpha G^\mathbf{k} d_\alpha G^\mathbf{k}$, for which
$\mathbf{g}=\mathbf{k} + d_\alpha \mathbf{k}$ is a vector of the reciprocal lattice.
This latter condition accounts for the Cooper pair's 
vanishing total momentum (modulo a reciprocal lattice vector). Introducing the 
intersection of wavevector groups
$M_\sigma = G^\mathbf{k} \cap d_\sigma G^\mathbf{k}d^{-1}_\sigma$, 
and choosing an element  
$a \in d_\sigma G^\mathbf{k} \cap G^\mathbf{k} d^{-1}_\sigma$,  
$P^-_\sigma$ is  then the representation of 
$\tilde{M}_\sigma= M_\sigma + a M_\sigma$
induced from $\gamma_\mathbf{k}$ by the following definition 
of its characters  
($m\in M_\sigma$)
\begin{align}
\label{eq:2}
\chi(P^-_\sigma(m)) =& \chi(\gamma_\mathbf{k}(m)) 
\chi(\gamma_\mathbf{k}(d^{-1}_\sigma m d_\sigma)), \\
\label{eq:3}
\chi(P^-_\sigma(a m)) =& -\chi(\gamma_\mathbf{k}(amam)). 
\end{align}

Turning to (ii), induced representations are conveniently 
calculated with the help of the `Frobenius reciprocity 
theorem'~\cite{brad72}. 
In the context of Eq.~(\ref{eq:1}) the theorem states 
that the number of times $n_j$ an irreducible representation $\Gamma^j$ 
of $G$  
appears in the decomposition ${\cal P}(\mathbf{k})=\sum_j n_j\Gamma^j$
equals the number of times $P^-_\sigma$ appears in
the decomposition of $\Gamma^j$
into irreducible representations $\tilde{\Gamma}^j$ 
of the `subduced' group $G\cap\tilde{M}$.
Here, both $\Gamma^j$ and $\tilde{\Gamma}^j$
are representations at the zone center.
We summarize $\tilde{\Gamma}^j$ for points of interest in Table~\ref{table:1}.
\begin{table}[h!]
\begin{tabular}{p{.4cm}|p{.8cm}p{.6cm}}
 $\tilde{\Gamma}^j$ & $(E,0)$ & $(I,0)$ \\
\hline 
$\Gamma_g$ & $\quad1$   & $\quad1$ \\
$\Gamma_u$ & $\quad1$   &  $\,\,-1$
\end{tabular}
\hspace{.9cm}
\begin{tabular}{p{.4cm}|p{.8cm}p{1cm}p{.7cm}p{.9cm}}
 $\tilde{\Gamma}^j$ & $(E,0)$ & $(2_z,\mathbf{t}_2)$ & $(I,0)$ & $(\sigma_z,\mathbf{t}_2)$\\
\hline 
$A_g$ & $\quad 1$   & $\quad 1$ & $\quad 1$ & $\quad 1$\\
$B_g$ & $\quad 1$   & $\,\, -1$ & $\quad1$ & $\,\,-1$\\
$A_u$ & $\quad 1$   & $\quad 1$ & $\,\,-1$ & $\,\,-1$\\
$B_u$ & $\quad 1$   & $\,\,-1$ & $\,\,-1$ & $\quad 1$
\end{tabular}
\caption{Characters of representations $\tilde{\Gamma}^j$ of the 
subduced groups. The left table applies to the group $G\cap\tilde{M}$ for a general point
in the zone; the right table to the group $G\cap\tilde{M}$ 
for points in the planes $k_z=0$ and $k_z=\pi/c$.
Here $g$ refers to even parity, and $u$ odd parity representations.}
\label{table:1}
\end{table}
Line nodes of the odd parity 
Cooper pair wavefunction 
may arise if any of the odd parity representations $\tilde{\Gamma}^j_{u}$ 
are absent in the decomposition
$P^-_\sigma=\sum_j m_j \tilde{\Gamma}^j$
in a symmetry plane of the zone
that intersects the Fermi surface.

To apply the outlined procedure, we need
to identify small representations $\gamma_{\mathbf{k}}$.
Our main focus is on $\mathbf{k}$ vectors on the
zone face $k_z=\pi/c$ (ZF). 
For purpose of illustration we also discuss 
the symmetry plane $k_z=0$ (SP) and a 
general $\mathbf{k}$ point (GP).  
In the presence of spin-orbit interactions, spin
and real space do not transform independently,
and the spin rotation group is absorbed into 
the crystal's space group. Moreover, 
extra degeneracies may occur due to time-reversal 
symmetry.
Both effects are taken into account when considering
co-representations of the little group 
$G^\mathbf{k}$. 

For illustration let us derive representations 
$P^-_\sigma$ for a general $\mathbf{k}$ point: 
the little group 
$G^{\rm GP}$ consists only of the identity and 
its multiplication with primitive translations. 
A co-representation $\gamma_{\rm GP}$ is
characterized by the  identity's character,
$\chi((E,0))$, which is two from
the fact that any point in the  zone
is twofold degenerate (Kramers theorem). 
The only double coset representative
satisfying the zero-momentum condition is 
$d_1=(I,0)$. $M_1$ is identical to $G^{GP}$ 
and $a=(I,0)$.  
The representation $P^-_1$ is
then readily deduced from Eqs.~(\ref{eq:2}) and
(\ref{eq:3}), see Table~\ref{table:2}. 
The decomposition of $P_1^-$ into representations
$\tilde{\Gamma}^j$ of  Table~\ref{table:1} results in
$P^-_1= \Gamma_g + 3\Gamma_u$.
This corresponds to Anderson's classification of the Cooper pair
wavefunction into an even parity pseudo-spin singlet and 
an odd parity pseudo-spin triplet. 
At a general $\mathbf{k}$ point, there is no symmetry reason for any of them to vanish.

Next, we turn to points in the symmetry planes: little 
groups $G^{\rm SP}$ and $G^{\rm ZF}$ are both formed by $(E,0)$,
$(\sigma_z,\mathbf{t}_2)$, and their multiplication 
with primitive translations. 
To account for
the appearance of non-trivial phase factors, one 
has to resort to the little groups'
central extensions and look at their projective 
representations~\cite{brad72}. 
One may readily convince oneself of the absence of
non-trivial phase factors for wavevectors that satisfy $k_z=0$.
 For points on the zone face, however, non-trivial
 phase factors arise. 
As a result, $G^{\rm SP}$
and $G^{\rm ZF}$ define different groups. They can be 
identified by their multiplication table, and 
their co-representations can be looked up. 
The (relevant) characters of the co-representations 
are as follows:  for $k_z=0$,
there are two identical co-representations $\gamma_{\rm SP}$ 
characterized by $\chi((E,0))=2$ and $\chi((\sigma_z,\mathbf{t}_2))=0$.
At the zone face, on the other hand, there are two complex 
conjugate co-representations $\gamma_{\rm ZF}$ with characters 
$\chi((E,0))=2$ and $\chi((\sigma_z,\mathbf{t}_2))=\pm 2i$. 
The different characters $\chi((\sigma_z,\mathbf{t}_2))$
for these two cases
reflect the different type of degeneracy encountered,
i.e. a pairing  degeneracy 
and a doubling degeneracy, respectively~\cite{Lax}.

\begin{table}[h!]
\begin{tabular}{p{.6cm}|p{.8cm}p{.7cm}}
 $P_1^-$ & $(E,0)$ & $(I,0)$\\
\hline 
GP & $\quad4$   & $-2$ 
\end{tabular}
\hspace{.9cm}
\begin{tabular}{p{.7cm}|p{.8cm}p{1cm}p{.7cm}p{.9cm}}
 $P_{1,2}^-$ & $(E,0)$ & $(2_z,\mathbf{t}_2)$ & $(I,0)$ & $(\sigma_z,\mathbf{t}_2)$\\
\hline 
SP & $\quad4$   & $\quad2$ & $-2$ & $\quad0$\\
ZF & $\quad4$ & $\,\,-2$ & $-2$& $\quad4$
 \end{tabular}
\caption{
Representations $P_i^-$ induced by $\gamma_\mathbf{k}$.
Left table: representation for a general $\mathbf{k}$ point.
Right table: representations for  
 $k_z=0$ and  
 $k_z=\pi/c$ are given by the first and second line, 
respectively.}
\label{table:2}
\end{table} 

Wavevectors for both symmetry planes allow 
for  $d_1=(I,0)$ and $d_2=(2_z,\mathbf{t}_2)$, resulting in 
$M_{1,2}$ both identical to the little group. Also, it is always possible
to choose $a=(I,0)$. Application of Eqs.~(\ref{eq:2}) and 
(\ref{eq:3}) to $k_z=0$ and 
$d_1$ results in the first line of the right 
table~\ref{table:2}~\cite{details1}. Using instead $d_2$ leads to 
the identical result in the second line (thus $P_1^- \equiv P_2^-$).
The decomposition into irreducible components $\tilde{\Gamma}^j$ 
of Table \ref{table:1} is
$P_i^- = A_g + 2A_u + B_u$, 
showing that half of the even parity representations ($B_g$) vanish for $k_z=0$. 
Odd parity representations, on the other hand, are all present,
indicating the absence of line nodes. This is
a consequence of Blount's argument, since phase factors 
related to the twofold screw axis are all trivial.

At the zone face, $d_1$ and 
$d_2$ lead again to identical representations. 
Also, 
results for both co-representations $\gamma_{ZF}$ (i.e. for 
characters $\chi((\sigma_z,\mathbf{t}_2))=\pm2i$)
are identical. The result is shown in the second line of the right 
Table~\ref{table:2}~\cite{details2}.
The decomposition
\begin{align}
\label{eq:4}
P_{1,2}^- = A_g + 3 B_u 
\end{align} implies that  half of the odd parity 
representations ($A_u$) vanish.
Eq.~(\ref{eq:4}) is the main result of this paper. 
It shows that in crystals with a twofold screw axis,
line nodes for odd parity Cooper pair wavefunctions
may occur whenever the Fermi surface intersects 
the Brillouin zone face perpendicular to the screw axis.
Our finding has relevance for a variety of 
unconventional superconductors.

{\it Implications:---}We first discuss the 
heavy fermion metal UPt$_3$, which was mentioned before.
Its non-symmorphic space 
group $P6_3/mmc$  possesses a twofold screw axis perpendicular to
the $k_z=\pi/c$ face of the hexagonal zone. 
Two of the Fermi surface sheets 
intersect this zone face~\cite{greg}. 
From our above analysis, it follows that
for $k_z=\pi/c$,
only those odd parity representations 
(of point group $6/mmm$) 
are allowed that are even under the operation $z\rightarrow -z$.
That is, an odd parity wavefunction belonging to the representations
$A_{1u}$, $A_{2u}$ or $E_{2u}$ has line nodes on the Fermi surface.
This potentially clears up a major puzzle in this material, where
various measurements are consistent with a line of nodes~\cite{upt3},
but the Knight shift indicates a spin triplet order parameter~\cite{tou}.
We note that an E$_{2u}$ order parameter has been proposed to
explain various experimental properties of UPt$_3$~\cite{e2u},
and recent phase sensitive measurements are in support of this
proposal~\cite{dvh}.

Another heavy fermion superconductor to which our observation
applies is UBe$_{13}$.  Again, the Knight shift suggests a spin triplet
state~\cite{tou},
while measurements of the NMR relaxation rate find a power law 
consistent with the presence of a line of nodes~\cite{amato}. 
UBe$_{13}$ has the 
non-symmorphic space group
$Fm{\bar 3}c$ that has twofold 
screw axes perpendicular to the square faces 
of the face centered cubic zone.
The Fermi surface is predicted to have pockets
that intersect these faces~\cite{fsUBe}.
Therefore, odd parity Cooper pair wavefunctions
belonging to the representations
(of point group $m\bar{3}m$)
$A_{1u}$, $A_{2u}$ or $E_{u}$
should have line nodes on the Fermi surface.

Our next example concerns the
recently discovered non-centrosymmetric 
superconductor Li$_2$Pt$_3$B~\cite{noncentrosym}.
Measurements of the temperature dependent
penetration depth point towards the existence of line nodes.
This finding has been 
attributed to a mixing of even and
odd parity components of the Cooper pair
wavefunction~\cite{yuan}. 
In crystals without inversion symmetry,
spin-orbit 
coupling lifts the Kramers degeneracy of the $\mathbf{k}$ states.
If the energy splitting $s$ resulting from this
is sufficiently large compared to the superconducting gap,
Cooper pairs can be admixed, $\Delta_\pm=\psi\pm t$,
with pseudo-spin singlet  
and triplet components, $\psi$ and $t$ respectively~\cite{singlettriplet}. 
If $t$ is large enough,
$\Delta_-$ may change sign, and thus a line of nodes is possible.
Given our above findings, we propose a second mechanism
for the appearance of a line of nodes in Li$_2$Pt$_3$B that would occur
in the opposite limit of weak
spin-orbit splitting of the bands.  Li$_2$Pt$_3$B has the space group symmetry
$P4_132$. This exhibits a twofold screw axis perpendicular to the
faces of the simple cubic zone.  The Fermi surface of Li$_2$Pt$_3$B 
is predicted to have several small pockets that intersect these faces~\cite{fsLiPtB}.
If $s$ is small enough that the mixing of odd and even parity components is not important, 
then Cooper pair wavefunctions belonging to the representations 
(of point group $m\bar{3}m$) 
$A_{1u}$, $A_{2u}$ or $E_u$
can have line nodes on the Fermi surface.
In contrast to the first scenario,
these line nodes are now constrained by symmetry.
Experiments should be able to differentiate
between these two scenarios.

Finally, we mention the more exotic example of  
Na$_4$Ir$_3$O$_8$, which is a candidate for a 
3D spin liquid~\cite{nairo}.
It has been proposed that this material possesses a `spinon'
Fermi surface~\cite{lawler,zhou}.
At the lowest temperatures, however,  the specific heat
decreases to zero as $T^2$, indicating (within this scenario) a line of nodes on this
spinon surface. This phenomenon has been recently 
attributed to pairing of the spinons in a mixed state as
described above for Li$_2$Pt$_3$B~\cite{zhou}.
Interestingly,  Na$_4$Ir$_3$O$_8$ has the same space group
P4$_1$32, and the predicted spinon Fermi surface also intersects the
simple cubic zone faces. Therefore, our previous discussion
for Li$_2$Pt$_3$B applies to this material as well, 
and we conclude that a pure triplet state with a line of nodes
is also possible.

{\it Conclusions:---}We have shown that 
in some non-symmorphic
superconductors,
it is possible to reconcile the existence of 
line nodes of an odd parity Cooper pair
wavefunction with the presence of (strong) spin-orbit
interactions. Specifically, we have proven that Blount's theorem
is superseded for
superconductors possessing a twofold screw axis
with a Fermi surface intersecting the zone face perpendicular to this axis. 
Our observation has potential relevance to a variety of unconventional
superconductors and spin liquids.

Work at Argonne National Laboratory was supported 
by the U. S. DOE, Office of Science, under Contract 
No.~DE-AC02-06CH11357.


\begin{thebibliography}{99}
\bibitem{example}
To be more specific, consider the eigenfunction 
of some space group element 
$(R,\mathbf{t})f_\mathbf{k}= cf_\mathbf{k}$
($R$ is a point-group
operation and $\mathbf{t}$ a translation vector).
The action of time-reversal symmetry $\theta$ and inversion symmetry 
$(I,0)$ generates new eigenfunctions with eigenvalues
$c$ and $e^{2i\mathbf{k}\mathbf{t}}c$, respectively, as can be checked 
by noting that $\theta$ and $(R,\mathbf{t})$
commute while 
$(R,\mathbf{t}) (I,0) =  (E, 2\mathbf{t}) (I,0) (R,\mathbf{t})$,
where $E$ is the point group identity.
It is then
evident that symmetries involving
non-primitive translations
generate different phase factors 
for the different 
pseudo-spin triplet components
at different points in the zone.
\bibitem{details1}
Notice that the characters are those of double groups,
and therefore for a $2\pi$ rotation,
$\chi((2_z2_z,0))=-1$.
\bibitem{details2}
Notice the additional factor $-1$  
for $\mathbf{k}$ points on the zone face
compared to $k_z=0$. This
arises from
$\chi((\gamma_\mathbf{k}(d_\sigma^{-1}(\sigma_z,\mathbf{t_2}) d_\sigma))
=\chi(\gamma_\mathbf{k}(\sigma_z,-\mathbf{t}_2))
=-\chi(\gamma_\mathbf{k}(\sigma_z,\mathbf{t}_2))$, i.e. 
the action of the translation group, 
$e^{2i\mathbf{k}\mathbf{t}_2}=-1$. 
Here $d_\sigma=(I,0)$ or $(2_z,\mathbf{t}_2)$ and we 
used the identity 
$(R,\mathbf{v})^{-1}(S,\mathbf{w})(R,\mathbf{v})
=(R^{-1}SR,R^{-1}S\mathbf{v}+R^{-1}\mathbf{w}-R^{-1}\mathbf{v})$.
\end{thebibliography}
\end{document}